\documentclass[journal]{vgtc}                
\ifpdf
  \pdfoutput=1\relax                   
  \usepackage{graphicx}                
  \DeclareGraphicsExtensions{.pdf,.png,.jpg,.jpeg} 
\else
  \usepackage{graphicx}                
  \DeclareGraphicsExtensions{.eps}     
\fi%

\graphicspath{{figures/}{pictures/}{images/}{./}} 

\usepackage{microtype}                 
\usepackage{hyperref}
\PassOptionsToPackage{warn}{textcomp}  
\usepackage{textcomp}                  
\usepackage{mathptmx}                  
\usepackage{times}                     
\usepackage{cite}                      
\usepackage{tabu}                      
\usepackage{booktabs}                  

\onlineid{0}

\vgtccategory{Research}
\vgtcpapertype{system}
\newcommand{\thename}{VisFCAC}
\title{\thename{}: An Interactive Family Clinical Attribute Comparison}

\author{Jake Gonzalez, Ngan V.T. Nguyen, and Tommy Dang}
\authorfooter{
\item
 The authors are with the Department
of Computer Science, Texas Tech University, Lubbock,
TX, USA. E-mail: \{jake.gonzalez\,$|$\,ngan.v.t.nguyen\,$|$\,tommy.dang\}@ttu.edu.
}

\shortauthortitle{Biv \MakeLowercase{\textit{et al.}}: Global Illumination for Fun and Profit}

\abstract{This paper presents \thename{}, a visual analysis system that displays family structures along with clinical attribute of family members to effectively uncover patterns related to suicide deaths for submission to the BioVis 2020 Data Challenge. \thename{} facilitates pattern tracing to offer insight on potential clinical attributes that might connect suicide deaths while also attempting to offer insight to prevent future suicides by at risk people with similar detected patterns. This paper lays out an approach to compare family members within a family structure to uncover patterns that may appear in clinical diagnosis data. This approach also compares two different families and their family structures to see whether there are patterns in suicide cases amongst clinical attributes outside family structures. Our solution implements a radial tree to display family structures with clinical attributes displayed on radial charts to provide in depth visual analysis and offer a comprehensive insight for underlying pattern discovery.} 


\CCScatlist{ 
 \CCScat{K.6.1}{Management of Computing and Information Systems}%
{Project and People Management}{Life Cycle};
 \CCScat{K.7.m}{The Computing Profession}{Miscellaneous}{Ethics}
}

\teaser{
  \centering
  \includegraphics[width=\linewidth]{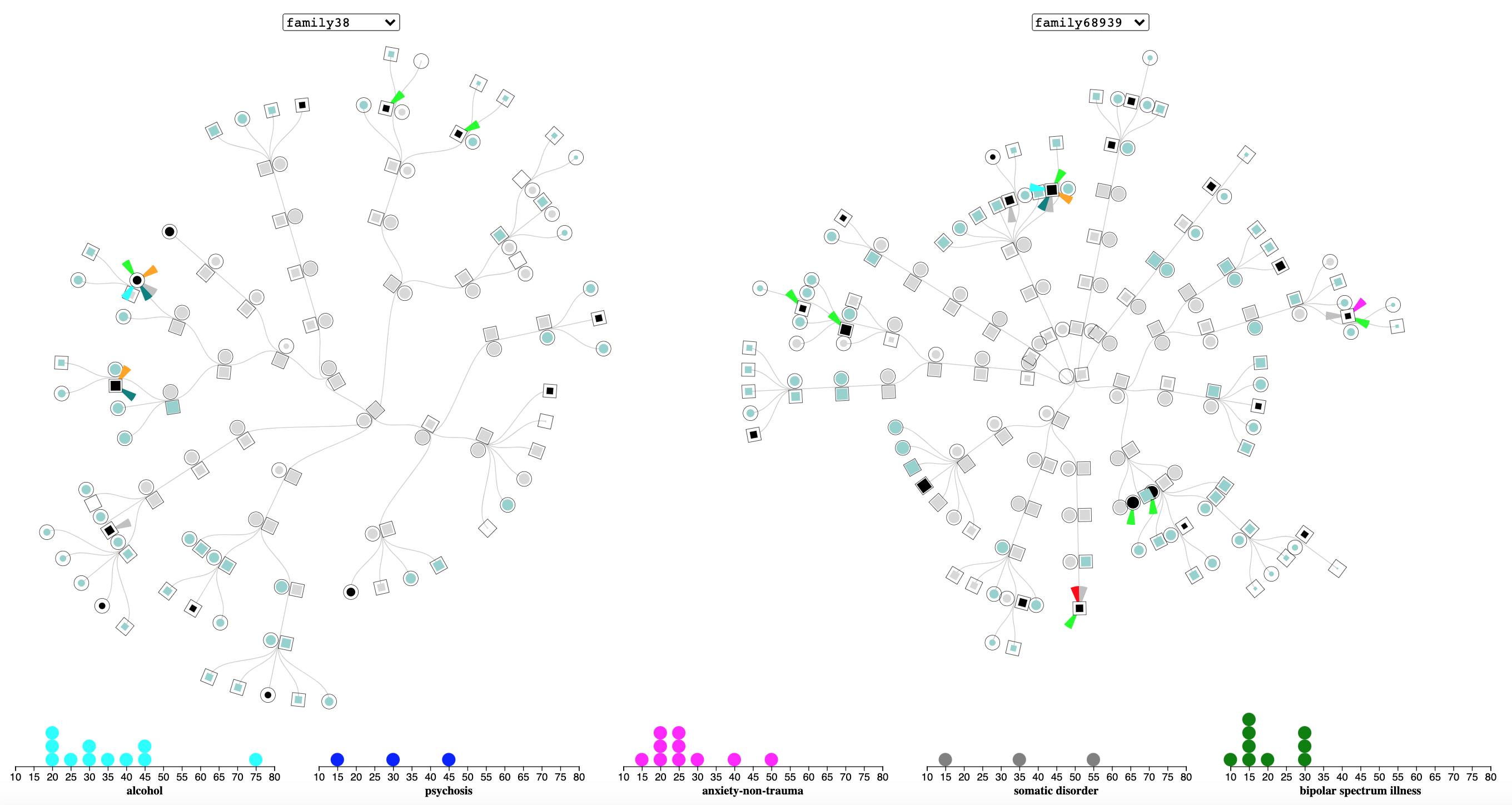}
  \caption{The \thename{} interface. The system presents the structure of the family tree that is selected. Squares represent males while circles represent females. The age of an individual is represented by the total area shaded in the nodes. Black shaded nodes represent individuals who have committed suicide, gray shaded nodes represent deceased individuals, and teal represents alive individuals. Family members that have committed suicide also contain a radial chart representing the individual's diagnosed clinical attributes. }
  \label{fig:teaser}
}
\begin{document}



\maketitle


\section{Introduction}
\label{sec:intro}
The BioVis 2020 Data Challenge focuses on public health data to uncover the underlying patterns of disease that may play a role in the rate of suicide. The challenge tasks participants to create a novel visualization to compare family structures with diagnosed clinical attributes of the family members to determine patterns that involve suicide cases. Because it is important to clearly show how hierarchical relationships influence other elements\cite{treeVis}, a radial tree visualization with generational separation is used to show the hierarchical structure of nine families from the Utah Population Database. To tackle this challenge, the structure of two family trees are displayed side-by-side for comparison. Family members who fell victim to suicide are represented on the tree by a radial chart showing the person's diagnosed clinical attributes.

We built a comprehensive web-based visual analysis system that allows users to compare two family trees and the family members that have fallen victim to suicide in order to find underlying patterns in diagnosed clinical attributes. The view of \thename{} is shown in Figure \ref{fig:teaser}. The general workflow includes 1) selecting a family to be displayed on the right, and one to be displayed on the left. Then 2) analyze family members that have fallen victim to suicide and analyze their diagnosed clinical attributes. 3) Compare the clinical attributes contained amongst family members who have committed suicide to uncover patterns of clinical attributes both within the family structure, and amongst other families. 4) Use the dot-plot of a clinical disease to determine the distribution of the disease amongst the families with the age of diagnosis to help uncover patterns. A similar analytical approach using a graph visualizations on social media data to uncover relationship patterns can be found in~\cite{visMCA}. The complete source code for the visualization can be found at \url{https://jakettu.github.io/BioVis2020/}.

\section{System Architecture}

\subsection{Radial Tree}
The view is presented in Figure \ref{fig:teaser}. This allows the user to select two of the nine families whose data was given in the challenge in order to compare the family hierarchical structure of the two families. Each family unit is represented as a square and a circle, the father and mother representatively. Or if the person is a child that is not found to be with a spouse, they are represented my an individual square or circle. The shaded area within the shape represents individual's age, the larger the shaded area, the older. The color is representative of whether the individual is alive, deceased, or has fallen victim to suicide. A teal represents the individual is alive, a gray represents the individual is deceased, and black represents a suicide. The root of the tree is representative of the first generation that was given in the data from the Utah Population Database, and each subsequent generation is located at the end of a link from the previous generation. The tree is also separated by generation.

\subsection{Radial Chart}
\begin{figure}[tb]
 \centering 
 \includegraphics[width=\columnwidth]{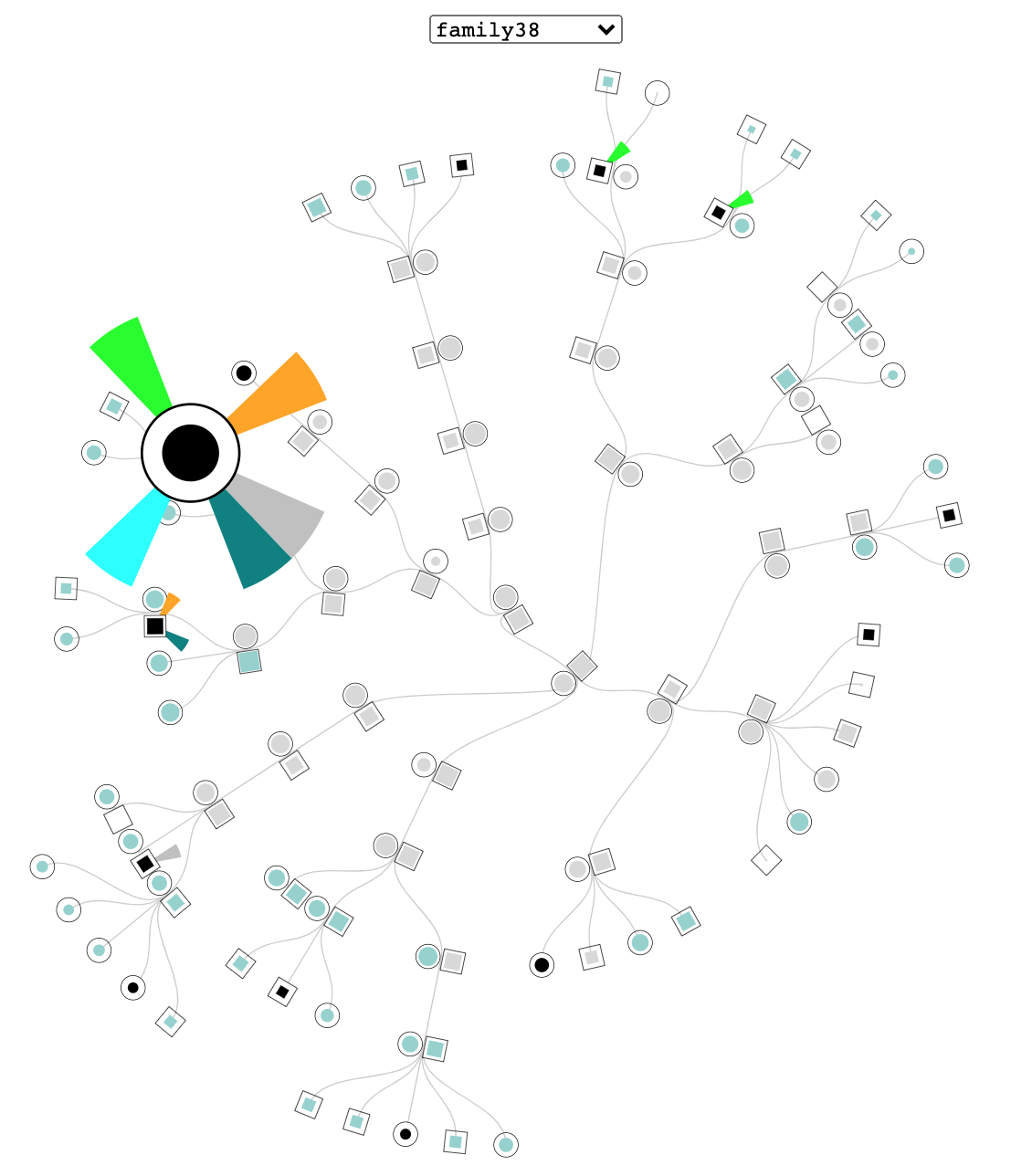}
 \caption{When analyzing the diagnosed clinical attributes of a family member who has committed suicide, mousing over the radial chart enlarges the graphic to more clearly see the colored portions which can be referenced with the color of the corresponding dot-plot at the bottom of the visualization to determine the diagnosed disease}
 \label{fig:nine}
\end{figure}

Figure \ref{fig:nine} displays the visualization used to investigate clinical attributes of a suicide victim. A user can simply hover the mouse over an individual who has committed suicide with clinical diagnosis data to enlarge the radial chart. The chart represents sixteen different clinical diagnoses.
Each of these clinical diagnoses are represented by a unique portion of the radial chart, and are represented by a unique color. The user can reference the dot-plots at the bottom of the visualization for the mapping of color to disease.

Using this platform, the user can analyze if there was a point in the family tree where a clinical disease appeared in the family, and which family members might have inherited genes that also make them susceptible to the same clinical disease. Also, if a clinical disease appears in different branches of the family tree, the user is able to easily determine who the last common ancestor was to offer a deeper insight.  

\subsection{Comparison}
The side-by-side comparison becomes useful when tracking clinical diagnoses within a family to determine if that clinical attribute follows a similar pattern amongst the other families in the given data set. Each family tree structure varies greatly, thus patterns in disease heredity can easily be uncovered. By comparing different family structures, some environmental factors can also be investigated to determine if they play a role suicide rates and clinical attributes. For example if a family member commits suicide, the likelihood of a family member later on in that family structure committing suicide can be evaluated. 

A dot-plot at the bottom of the visualization can be used to determine the distribution of suicide victims' diagnosed clinical attributes. The dot-plot shows the total number of individuals that were diagnosed with a particular disease. When mousing over the circles on the dot-plot, the user can identify the person id of the individual, the family the individual belongs to, and the age of diagnosis.  

\section{Challenge Findings}
With the aid of the created visualization, we can determine that depression is one clinical attribute amongst family members who have fallen victim to suicide and can easily be passed from one generation to the next. Even if a person in the previous generation does not contain data to definitively determine if they suffered from depression, we can see that there may very well be a gene that predisposes offspring to suffer from this disease. We can see in families 27251 and 68939 that there are different family members of the same generation that have fallen victim to suicide where each of them were diagnosed with depression. This may be caused by biological factors, or may be caused by environmental factors. In the case of family 27251, we can see that both offspring of a suicide victim were diagnosed with depression, and then also fell victim to suicide. 

While it is difficult for us to determine how the diagnosed individuals came to have these diseases, whether being biological or environmental factors, the user can easily uncover patterns of clinical diagnoses and suicide to know where to focus deeper investigation. It can also potentially offer some insight on why some members of a family within a generation seem healthy with few deaths, but have one family member who committed suicide and suffered from numerous clinical diseases. For example in family 149, in the ninth generation a family member was diagnosed with 5 different clinical diseases before committing suicide, while all members of that generation, the previous generation, and the second previous generation are still alive. 

\section{Conclusion}
This paper presents \thename{}, a visual analysis system implemented in HTML5 with D3.js for uncovering patterns in suicide rates amongst different family structures. The system enables a quick overview of the structure of the nine different families in the Utah Population Database and shows which of the family members are alive, deceased, and have fallen victim to suicide. With investigation by the user, the radial chart offers insight into why patterns of suicide cases may exist within a family structure. While it difficult to determine the underlying cause of clinical diseases, patterns of illness and suicide can be traced to determine if family members of the current or future generations may be at risk to suicide or illness. With the side-by-side comparison view of different family structures, the user can easily determine what clinical diseases are more prone to inheritability by comparing how a given clinical disease forms patterns in different family structures.


\bibliographystyle{abbrv-doi}

\bibliography{template}
\end{document}